\documentclass[prl,aps,twocolumn,longbibliography,superscriptaddress]{revtex4-1}
\usepackage{graphicx,amsmath,bm}
\usepackage{txfonts}
\usepackage{color}
\usepackage{mathdots}
\usepackage{enumitem}
\usepackage{bibunits}
\usepackage{hyperref}
\usepackage{notes2bib}
\bibnotesetup{
use-sort-key = false
}
\newcommand{\citeN}[1]{\bibnotemark[#1]}

\newcommand{\rp}[1]{(\ref{#1})}

\newcommand{\abs}[1]{\left|{#1}\right|}

\newcommand{\av}[1]{\left\langle #1 \right\rangle}

\newcommand{\br}[1]{\langle #1|}

\newcommand{\ke}[1]{|#1\rangle}

\newcommand{\wt}[0]{\widetilde}

\newcommand{\al}[1]{^{(#1)}}
\newcommand{\da}{^\dagger}

\newcommand{\pt}[1]{\left( #1 \right)}
\newcommand{\pq}[1]{\left[ #1 \right]}
\newcommand{\pg}[1]{\left\{ #1 \right\}}

\newcommand{\lpg}[1]{\left\{ #1 \right.}

\newcommand{\rpg}[1]{\left. #1 \right\}}
\newcommand{\ee}{{\rm e}}
\newcommand{\ii}{{\rm i}}

\newcommand{\id}{\openone}

\newcommand{\nn}{{\nonumber}}

\newcommand{\mat}[2]{
                      \begin{array}{#1}
                       #2
                       \end{array}  }

\newcommand{\matt}[2]{ \pt{
                      \begin{array}{cc}
                       #1 \\
                       #2
                     \end{array}  }  }

\newcommand{\vb}{{\bf b}}

\newcommand{\vv}{{\bf v}}
\newcommand{\vw}{{\bf w}}
\newcommand{\vx}{{\bf x}}

\newcommand{\AAA}{{\cal A}}
\newcommand{\BB}{{\cal B}}
\newcommand{\CC}{{\cal C}}
\newcommand{\DD}{{\cal D}}
\newcommand{\GG}{{\cal G}}

\newcommand{\JJ}{{\cal J}}
\newcommand{\KK}{{\cal K}}
\newcommand{\II}{{\cal I}}
\newcommand{\LL}{{\cal L}}
\newcommand{\MM}{{\cal M}}

\newcommand{\OO}{{\cal O}}

\newcommand{\QQ}{{\cal Q}}
\newcommand{\RR}{{\cal R}}
\newcommand{\TT}{{\cal T}}
\newcommand{\UU}{{\cal U}}
\newcommand{\VV}{{\cal V}}
\newcommand{\WW}{{\cal W}}
\newcommand{\XX}{{\cal X}}
\newcommand{\YY}{{\cal Y}}
\newcommand{\ZZ}{{\cal Z}}
\newcommand{\SSS}{{\cal S}}

\definecolor{mygreen}{rgb}{0., 0.6, 0.}

\usepackage[normalem]{ulem}

\begin{document}

\title{
Dissipative engineering of Gaussian entangled states in harmonic lattices with a single-site squeezed reservoir
}

\author{Stefano Zippilli}
\affiliation{School of Science and Technology, Physics Division, University of Camerino, I-62032 Camerino (MC), Italy}
\author{David Vitali}
\affiliation{School of Science and Technology, Physics Division, University of Camerino, I-62032 Camerino (MC), Italy}
\affiliation{INFN, Sezione di Perugia, I-06123 Perugia, Italy}
\affiliation{CNR-INO, L.go Enrico Fermi 6, I-50125 Firenze, Italy}
\date{\today}

\begin{abstract}
We study the dissipative preparation of many-body entangled Gaussian states in bosonic lattice models which could be relevant for quantum technology applications. We assume minimal resources, represented by systems described by particle-conserving quadratic Hamiltonians, with a single localized squeezed reservoir. We show that in this way it is possible to prepare, in the steady state, the wide class of pure states which can be generated by applying a generic passive Gaussian transformation on a set of equally squeezed modes. This includes non-trivial multipartite entangled states such as cluster states suitable for measurement-based quantum computation.
\end{abstract}
\maketitle

\begin{bibunit}

The harnessing of quantum many-body dynamics by engineered dissipation
is interesting for applications in quantum technology~\cite{kraus2008,diehl2008,verstraete2009}.
In these approaches the environment of many interacting quantum systems is designed in such a way that the interplay between controlled dissipation and interactions results in specific controlled system dynamics~\cite{verstraete2009,kastoryano2013,zanardi2014,gong2017}, in the simulation of complex quantum system~\cite{weimer2010,barreiro2011,stannigel2014}, 
and in the robust preparation of non-trivial quantum global stationary states~\cite{kraus2008,diehl2008,diehl2011,cho2011a,morigi2015a,reiter2016}, including Gaussian states~\cite{koga2012}.
In general, the practical realization of these dynamics is hampered by the need to engineer the environment of all the many elements which constitute the system. 
However, it has been also shown that under certain conditions it is possible to engineer a single localized reservoir to have control over the global properties of the system~\cite{barontini2013,tonielli2019a,zippilli2013}.

In this work we are interested in strategies which make use of minimal
resources, namely only one squeezed reservoir and a bosonic lattice with a
passive (particle-conserving) quadratic Hamiltonian~\cite{zippilli2015,asjad2016a,ma2017,
yanay2018,yanay2020b,yanay2020a,zippilli2013,zippilli2014,ma2019}.
It has been shown that these systems can be steered into peculiar entangled steady states, when 
the squeezed reservoir is coupled to single site of the lattice and the Hamiltonian is endowed with
specific symmetries~\cite{zippilli2015,yanay2018}.
Here we characterize the class of Gaussian pure states that can be achieved with this approach, and we show that it is composed of all the states that can be generated by applying any combination of 
particle-conserving quadratic operations (beam splitters and phase shifts) on a set of equally squeezed modes. 
We also identify the general properties of the Hamiltonians which enable the generation of these pure stationary states (showing, in particular,  that they necessarily satisfy the chiral symmetry identified in Ref.~\cite{yanay2018}), and, for each state, we discuss how to construct the specific Hamiltonian which sustain such state in the stationary regime.
Interestingly, the class of states that can be obtained in this way includes Gaussian cluster states usable for universal measurement-based quantum computation with continuous variables~\cite{menicucci2006,gu2009}, and, as a prominent example, we study the performance of the present approach for the preparation of a cluster state in a square lattice.
In measurement-based quantum computation a big part of the complexity of the computation is placed into the preparation of the cluster state. In particular, optical setups are very promising and scalable platforms for this task~\cite{menicucci2007,
menicucci2008,flammia2009,menicucci2010,menicucci2011a,chen2014,yokoyama2013,
alexander2016a,cai2017a,alexander2018,su2018,larsen2019,wu2020a,asavanant2019,asavanant2020}. Our proposal suggests that similar results could be achieved also with localized quantum modes in, for example, circuit QED systems~\cite{hacohen-gourgy2015,fitzpatrick2017,ma2019b}.

In detail, we study the dissipative preparation of a zero-average pure Gaussian state of $N$ bosonic modes $\ke{\Psi}$, considering $N+1$ bosonic modes (including an additional auxiliary mode). 
They are described by the annihilation operators $b_j$ for $j\in\pg{0,1\cdots N}$, and we assume that only the auxiliary mode, that is the one with index $j=0$, is coupled to a squeezed reservoir.
In the ideal situation the auxiliary mode is the only open mode which is subject to dissipation in the squeezed reservoir. Additional dissipation acting on the other modes reduces the purity of the final state and will be addressed later on.
We assume quadratic Hamiltonians $H$ for the $N+1$ modes, with only passive interaction terms, $H=\hbar\,\sum_{j,k=0}^N\,\JJ_{j,k}\,b_j\da\,b_k$ (with $\JJ_{j,k}=\JJ_{k,j}^*$), which conserves the number of excitations, so that the existing quantum correlations in the steady state are a consequence of the correlations in the reservoirs only.
The system is described by the master equation
\begin{eqnarray}\label{MEq}
\dot\rho=-\frac{\ii}{\hbar}\pq{H,\rho}+\LL\,\rho,
\end{eqnarray}
where the effect of the squeezed bath is given by the Lindblad term 
$\LL\,\rho=\kappa\{
(\bar n+1)\,\DD_{b_0,b_0\da}+\bar n\,\DD_{b_0\da,b_0}
-\bar m^*\,
\DD_{b_0,b_0}-\bar m\,\DD_{b_0\da,b_0\da}
\}\rho$
with 
$\DD_{x,y}\ \rho=2\ x\ \rho\ y-y\ x\ \rho-\rho\ y\ x$,
and $\abs{\bar m}=\sqrt{\bar n(\bar n+1)}$ (this condition corresponds to a reservoir in a pure squeezed state; if $\abs{\bar m}<\sqrt{\bar n(\bar n+1)}$ the reservoir is not pure, and the states that we discuss here are modified, in a straightforward way, by a thermal component~\cite{zippilli2015}).
The central result of this work is the following theorem.
\paragraph*{Theorem.}
%
A zero-average pure Gaussian state which is factorized between the auxiliary mode ($\ke{\psi_0}$) and the remaining $N$ modes ($\ke{\Psi}$)
\begin{eqnarray}\label{Psitot}
\ke{\Psi_{tot}}=\ke{\psi_0}\,\ke{\Psi}\ ,
\end{eqnarray}
and is generated from the vacuum $|0\rangle $ by the unitary transformations $U_0$ and $U$,
such that $\ke{\psi_0}=U_0\,\ke{0}$ and $\ke{\Psi}=U\,\ke{0}$, is the unique steady state of Eq.~\rp{MEq} if and only if the following three propositions are true:
\begin{enumerate}[leftmargin=15pt]
\item \label{cond0}
	$U_0$ is the squeezing transformation 
	$U_0=\ee^{\frac{z_0}{2}\pt{\ee^{\ii\varphi_0}{b_0\da}^2-\ee^{-\ii\varphi_0}{b_0}^2}}$,
	where the squeezing strength $z_0$ and the squeezing phase $\varphi_0$ are determined by the squeezing of 
	the reservoir according to the relations $\tanh(z_0)=\sqrt{\bar n/(\bar n+1)}$,  
	and $\ee^{\ii\varphi_0}=\bar m/\abs{\bar m}$;
\item \label{cond1}
	$U$ can be decomposed as $U=U\al{p}\,U\al{S}$, where $U\al{S}$ is the product of 
	$N$ single-mode squeezing transformations with squeezing strength 
	equal to that of the transformation $U_0$, i.e. $U\al{S}=U_1\cdots\,U_N$, with
	$U_j=\ee^{\frac{z_0}{2}\pt{\ee^{\ii\varphi_j}{b_j\da}^2-\ee^{-\ii\varphi_j}{b_j}^2}}$,
	and $U\al{p}$ is a passive quadratic transformation
	(note that both $U\al{S}$ and $U\al{p}$ don't operate on the auxiliary mode);
\item \label{cond2}
	the passive quadratic Hamiltonian $H$ for the $N+1$ modes of Eq.~\rp{MEq} is given by
	$H=U\al{p}\ H\al{S}\ {U\al{p}}\da$,	where $H\al{S}$ is any passive quadratic Hamiltonian 
	for which the following propositions are true:
	\begin{enumerate}[leftmargin=16pt]
	\item \label{cond2a}
	$H\al{S}$ remains passive under the effect of the set of single-mode squeezing transformations 
	for the $N+1$ modes $U_0\,U\al{S}$, i.e. ${U\al{S}}\da\,U_0\da\ H\al{S}\ U_0\,U\al{S}$ is passive;
	\item \label{cond2b}
		all the normal modes of $H\al{S}$ have a finite overlap with the auxiliary mode (see~\cite{SM}).
	
	\end{enumerate}
\end{enumerate}

\paragraph*{Proof.}
%
\textit{Part 1: If the propositions~\ref{cond0}-\ref{cond2} are true  then Eq.~\rp{Psitot} is the only steady state.}
In the representation defined by the transformation $U_0\,U$, the transformed density matrix $\tilde\rho=U\da\,U_0\da\,\rho\,U_0\,U$, fulfills the master equation $\dot{\tilde\rho}=-\frac{\ii}{\hbar}\pq{\wt H,\tilde\rho}+\wt\LL\ \tilde\rho$ where the dissipative term, $\wt\LL\ \tilde\rho=\kappa\ \DD_{b_0,b_0\da}\ \tilde\rho$, describes pure dissipation in a vacuum reservoir, and the transformed Hamiltonian $\wt H=U\da\,U_0\da\,H\,U_0\,U$, can be written as $\wt H={U\al{S}}\da\,U_0\da\,H\al{S}\,U_0\,U\al{S}$. This shows that $\wt H$ is passive because of proposition~\ref{cond2a}.
The proposition~\ref{cond2b}, instead, entails that $H\al{S}$, and therefore also $H$ and $\wt H$, have no dark modes~\cite{SM}, i.e. all the normal modes are coupled to the reservoir. Thus, the only steady state in the new representation is the vacuum, which is equal to Eq.~\rp{Psitot} in the original representation.

\textit{
Part 2: If Eq.~\rp{Psitot} is the only steady state, then the propositions~\ref{cond0}-\ref{cond2} are true.}
In the representation defined by the density matrix $\tilde\rho$, the transformed steady state, $\ke{\wt\Psi_{tot}}=U\da\,U_0\da\,\ke{\Psi_{tot}}=\ke{0}$, is the vacuum. This can be true only if the transformed Hamiltonian $\wt H$ is passive with no dark modes, and the dissipative term $\wt\LL\ \tilde\rho=U_0\da\,\pq{\LL\,\pt{U_0\,\tilde\rho\,U_0\da} }U_0$ describes pure dissipation in a vacuum reservoir. For this to be true $U_0$ 
has to fulfills the proposition~\ref{cond0}.  

Now, in order to demonstrate the validity of the other propositions, we note that it is always possible to decompose a unitary transformation $U$, which generates a zero-average pure Gaussian state, in a form similar to the one defined in the proposition~\ref{cond1}, where $U\al{S}$ is a set of single-mode squeezing transformations which can be, in general, of different strength, and $U\al{p}$ is a multi-mode passive transformation. This can be seen by using the Bloch-Messiah decomposition~\citeN{SM}. Thus, Eq.~\rp{Psitot} can be always written in the form  $\ke{\Psi_{tot}}=U_0\,U\al{p}\,U\al{S}\,\ke{0}$.
In the representation defined by the transformed density matrix $\rho\al{S}={U\al{p}}\da\,\rho\,U\al{p}$, which fulfill the equation $\dot\rho\al{S}=-\frac{\ii}{\hbar}\pq{H\al{S},\rho\al{S}}+\LL\,\rho\al{S}$,
the Hamiltonian $H\al{S}={U\al{p}}\da\,H\,U\al{p}$ is passive (because $U\al{p}$ and $H$ are passive), and remains passive under the effect of $U_0\,U\al{S}$ (in fact ${U\al{S}}\da\,U_0\da\ H\al{S}\ U_0\,U\al{S}=\wt H$ which, as we have seen, has to be passive), and therefore the proposition~\ref{cond2a} is true.
Moreover, $\wt H$ has no dark modes (because we are assuming that the system has a single steady state), and thus the proposition~\ref{cond2b} is true as well~\cite{SM}.
Finally, this also means that all the modes are connected (even if not directly) by the interactions terms of $H\al{S}$, and this together with the following lemma guarantees that the strength of all the squeezing transformations which constitute $U\al{S}$ are equal.
In particular they have to be equal to the squeezing strength of the auxiliary mode $z_0$, which is fixed by the squeezing strengths of the reservoir, so also the proposition~\ref{cond1} is true.
$\hfill\blacksquare$

Let us now introduce the following lemma which describes the precise structure of the Hamiltonian $H\al{S}$.
\paragraph*{Lemma.}
Given a passive quadratic Hamiltonian, $H\al{S}=\hbar\,\sum_{j,k=0}^N\,\JJ_{j,k}\al{S}\,b_j\da\,b_k$, with $\JJ_{j,k}\al{S}=\abs{\JJ_{j,k}\al{S}}\,\ee^{\ii\,\Theta_{j,k}}$ and $\Theta_{j,k}=-\Theta_{k,j}$,
the transformed Hamiltonian $\wt H=U_N\da\,\cdots U_0\da\ H\al{S}\ U_0\cdots U_N$, with $U_j=\ee^{\frac{z_j}{2}\pt{\ee^{\ii\varphi_j}{b_j\da}^2-\ee^{-\ii\varphi_j}{b_j}^2}}$, is passive, if and only if
(i)
$\JJ_{j,j}\al{S}=0$ for all $j$ with $z_j\neq 0$, 
(ii)
$\Theta_{j,k}=n\,\pi+\pt{\varphi_j-\varphi_k+\pi}/{2}$ for $j<k$ (with $n\in\mathbb{Z}$), and $z_j=z_k$ for all $j\neq k$ with $\JJ_{j,k}\al{S}\neq 0$. Moreover, if $\wt H$  is passive then $\wt H=H\al{S}$.
(The  proof of this lemma is straightforward and is reported in~\cite{SM}).

\begin{figure*}[t!]
\centering
\includegraphics[width=\textwidth]{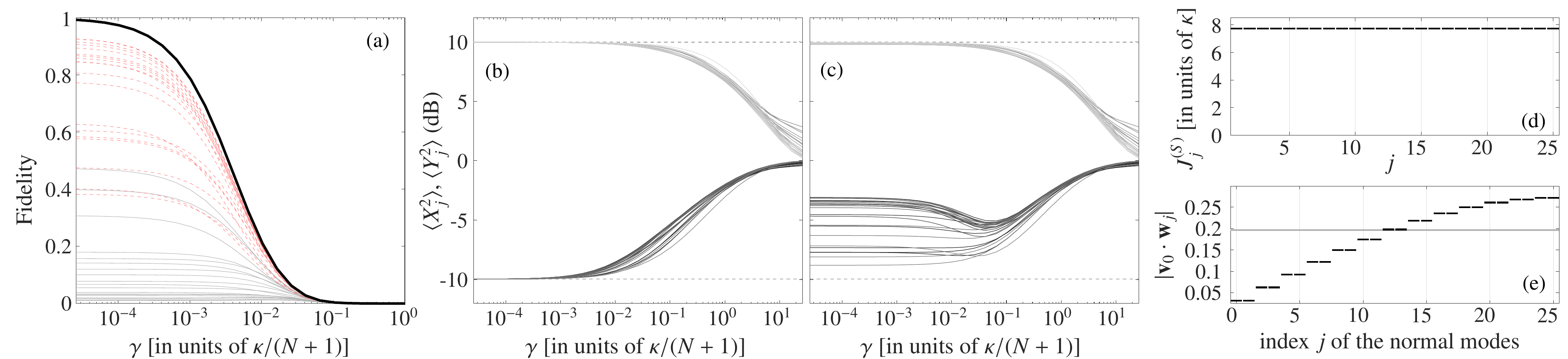}
\caption{Dissipative preparation of a cluster state of $N=25$ modes in a $5\times 5$ square lattice~\citeN{SM}. (a) Fidelity
${\rm Tr}\pg{\rho'_{st}\ \rho_{st}}$   
~\cite{spedalieri2013} between the steady state $\rho'_{st}$ of the model~\rp{MEqtot} and the corresponding steady state $\rho_{st}$ of Eq.~\rp{MEq}. The thick black line is evaluated using the Hamiltonian~\rp{H}
(with $J_j\al{S}=7.7\,\kappa$ [see panel (d)] and $\theta_j=0$, $\forall\,j$); 
the thin solid gray lines are evaluated for $20$ random realizations of the system Hamiltonian with interaction coefficients $\JJ_{j,k}\al{\zeta}=\JJ_{j,k}(1+\zeta_{j,k})$ where $\JJ_{jk}$ are the coefficients of $H$~\cite{SM}, and $\zeta_{j,k}$ are random variables uniformly distributed in the range $\pq{-0.001,0.001}$;
The thin dashed red lines are evaluated for $20$ random realizations of the system Hamiltonian with
$\JJ_{j,k}\al{\beta}=\JJ_{j,k}\,\ee^{\ii\,\beta_{j,k}}$ where $\beta_{j,k}=-\beta_{k,j}$ are random variables uniformly distributed in the range $\pq{-0.015,0.015}$.
(b), (c) Corresponding steady state variance of the normalized nullifiers $X_j=r_j\,x_j$ (lower dark gray lines) and of the orthogonal collective quadratures $Y_j=r_j\,y_j$  with all the modes rotated by $\pi/2$, such that $y_j=-q_j-\sum_{k=1}^N\,\AAA_{j,k}\,p_k$ (upper light gray lines), and where the normalization coefficients $r_j$ are chosen such that $X_j$ and $Y_j$ fulfill the standard commutation relation $\pq{X_j,Y_j}=2\,i$. Panel (b) corresponds to the thick black line of (a). Panel (c) corresponds to the realization (thin gray line) with 
the lowest fidelity of panel (a).
The horizontal dashed lines in (b) and (c) indicate the variance of the squeezed and anti-squeezed quadratures of the squeezed reservoir, which corresponds to $\bar n=2$.
(d) Interaction coefficients $J_j\al{S}$ of Eq.~\rp{HSlinearChain} used to compute the Hamiltonian~\rp{H}. (e) Corresponding overlap of the normal modes of $H\al{S}$ and the auxiliary mode, i.e. scalar product $|\vv_0\cdot\vw_j|$ between the normalized eigenvectors $\vw_j$ of the coefficient matrix $\JJ\al{S}$ of the Hamiltonian $H\al{S}$, and the vector, corresponding to the auxiliary mode, $\vv_0=\pt{1,0\cdots,0}$. The horizontal gray line in (e) indicates the value $1/\sqrt{N+1}$.}
\label{Fig1}
\end{figure*}

It is, now, important to point out that, for any given state $\ke{\Psi}$ which fulfills the proposition~\ref{cond1}, each quadratic Hamiltonian $H\al{S}$ which fulfills the propositions~\ref{cond2a}-\ref{cond2b} (and the lemma) can be used to construct a (different) Hamiltonian $H$ (see the proposition~\ref{cond2}) of model~\rp{MEq} which sustain the given state in the stationary regime.
Thus the same steady state can be obtained with many different Hamiltonians. The specific form of $H$ can determine how fast (and therefore how efficiently, when additional noise sources affect the system dynamics) the system approaches the steady state. 
We also note that both $H\al{S}$ and $H$ satisfy the chiral symmetry identified in Ref.~\cite{yanay2018} (see~\cite{SM}). 
This implies that the chiral symmetry of $H$, is also a necessary condition (not only a sufficient one, as suggested in Ref.~\cite{yanay2018}) for the existence of the pure steady state~\rp{Psitot} of Eq.~\rp{MEq}.

A particularly simple Hamiltonian $H\al{S}$ that fulfills the propositions~\ref{cond2a}-\ref{cond2b} (and the lemma) is the Hamiltonian for a linear chain with open boundary conditions (for which the normal modes have always a finite overlap with the end modes)
\begin{eqnarray}\label{HSlinearChain}
H\al{S}=\ii\,\hbar\,\sum_{j=1}^N\ J_j\al{S}\pt{\ee^{\ii\,\theta_j}\,b_{j-1}\,b_j\da-\ee^{-\ii\,\theta_j}\,b_{j-1}\da\,b_j}\ ,
\end{eqnarray}
where $\theta_j=\pt{\varphi_j-\varphi_{j-1}}/2$, with $\varphi_j$ the squeezing phases 
introduced in the proposition~\ref{cond1}. 
This means that Eq.~\rp{HSlinearChain} can be used to construct the Hamiltonian $H$ corresponding to any state that fulfills the proposition~\ref{cond1}.
Specific examples of multi-mode entangled states that can be prepared with this strategy have been  discussed in Ref.~\cite{zippilli2015,asjad2016a,ma2017,yanay2018,yanay2020b,yanay2020a}.

It is interesting to note that the class of states that can be prepared with our approach is wide and it includes also cluster states which are the main resource of measurement-based quantum computation~\cite{menicucci2006,gu2009}.
In particular all the cluster states that have been proposed and prepared by manipulating one or two squeezed light beams with a complex interferometer~\cite{menicucci2007,
menicucci2008,flammia2009,menicucci2010,menicucci2011a,chen2014,yokoyama2013,
alexander2016a,cai2017a,alexander2018,su2018,larsen2019,wu2020a,asavanant2019,asavanant2020} 
can be also generated following our approach.
The difference between these results and the present approach is that, while in these works the state is prepared in traveling wave beams of light, our results shows how to generate similar states, in a robust way, as stationary states of a dissipative dynamics. This approach is, hence, attractive in situations in which the quantum modes are localized, as for example in a solid-state or atomic device~\cite{ozawa2019,tomza2019}.

\begin{figure*}[t!]
\centering
\includegraphics[width=\textwidth]{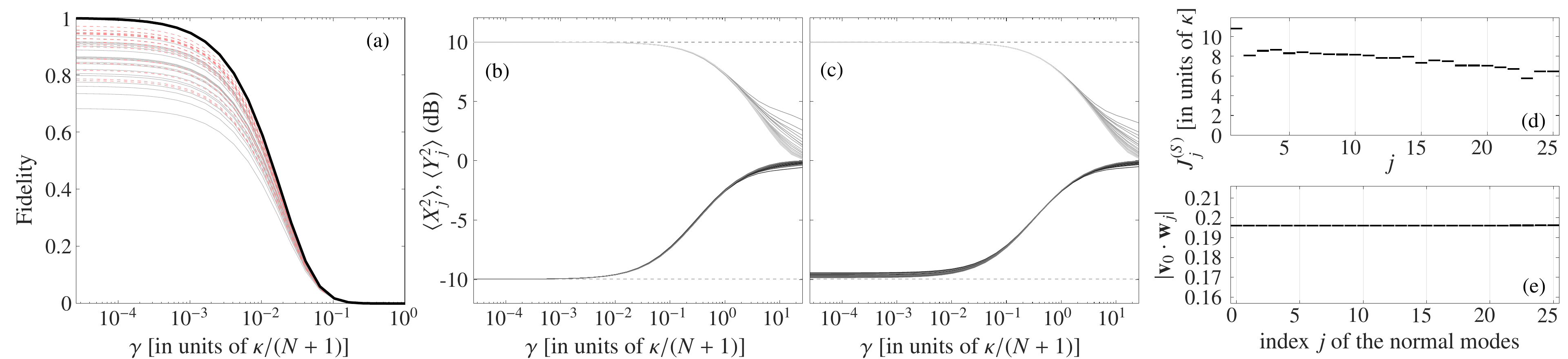}
\caption{As in Fig.~\ref{Fig1} with the values of the interaction coefficients $J_j\al{S}$ of $H\al{S}$~\rp{HSlinearChain} reported in panel (d) (note that the average value of these coefficients is equal to the  value of $J_j\al{S}$ used in Fig.~\ref{Fig1}). These coefficients have been found by the numerical maximization of the smallest overlap between the normal modes and the auxiliary mode, such that the resulting overlaps are all equal to $1/\sqrt{N+1}$ [see panel (e)].}
\label{Fig2}
\end{figure*}

\paragraph*{Dissipative generation of a cluster state.}
Let us now investigate the potentiality of our result to design a model which sustain in the stationary regime a cluster state in a square lattice~\citeN{SM} which constitutes a universal resource for measurement-based quantum computation~\cite{gu2009,su2018}.
To be specific, we consider a cluster state of $N=25$ modes with a $N\times N$ real symmetric adjacency matrix $\AAA$ (with non-zero entries equal to one) which represents the square lattice~\citeN{SM}. This state can be generated by the multi-mode squeezing transformation~\cite{zippilli2020}
$U_z=\ee^{-\ii\,\frac{z}{2}\sum_{j,k=1}^N\pt{\ZZ_{j,k}b_j\da\,b_k\da+\ZZ_{j,k}^*\,b_j\,b_k}}$,
where the $N\times N$ matrix of interaction coefficients is given by $\ZZ=-\ii\ \pt{\AAA-\ii\,\id}\,\pt{\AAA+\ii\,\id}^{-1}$.
What characterizes this as cluster state is the fact that
the covariance matrix of the $N$ operators $x_j=p_j-\sum_{k=1}^N\,\AAA_{j,k}\,q_k$ [with $q_j=b_j+b_j\da$ and $p_j=-\ii\pt{b_j-b_j\da}$], called nullifiers, approaches the null matrix in the limit of infinite squeezing, $z\to\infty$~\cite{zippilli2020}.
The transformation $U_z$ can be decomposed, similarly to the definition in the proposition~\ref{cond1} of the theorem, as $U_z=U_z\al{p}\,U_z\al{S}$, with $U\al{S}_z$ given by the product of $N$ equal single-mode squeezing transformations
(where $\varphi_j=0$ for all $j$), and with $U_z\al{p}$ which fulfills the relation ${U_z\al{p}}\da b_j\,U_z\al{p}=\sum_{k=1}^N\pg{\pt{-\ii\,\ZZ}^{1/2}}_{j,k}b_k$~\citeN{SM}.
The fact that $U_z\al{S}$ describes the equal squeezing of all the modes implies, according to our theorem, that 
$U_z\ke{0}$ is the steady state of Eq.~\rp{MEq} when 
\begin{eqnarray}\label{H}
H=U_z\al{p}\,H\al{S}\,{U_z\al{p}}\da\ ,
\end{eqnarray}
where $H\al{S}$ is the Hamiltonian for the linear chain~\rp{HSlinearChain}.
Note that the same cluster state, given by a specific adjacency matrix, can be generated by many different transformation $U_z$, which correspond to different $U_z\al{p}$~\cite{SM,zippilli2020,ferrini2015}, and thus to different $H$. 
The specific form of $H$ can be relevant and should be taken into account when considering an experimental implementation of these results.

In Fig.~\ref{Fig1} and \ref{Fig2} we show the results for the preparation of this cluster state.
We have studied how the present approach performs in non-ideal situations which include additional noise sources, with dissipation rate $\gamma$, and random deviations from the optimal system Hamiltonian defined in Eq.~\rp{H}. In particular, in Fig.~\ref{Fig1} and \ref{Fig2}, we characterize the steady state
$\rho'_{st}$ of
\begin{eqnarray}\label{MEqtot}
\dot\rho'=-\frac{\ii}{\hbar}\pq{H,\rho'}+\LL\,\rho'+\gamma\,\sum_{j=0}^N\,\DD_{b_j,b_j\da}\,\rho'\ ,
\end{eqnarray}
in terms of its fidelity with respect to the steady state $\rho_{st}$ achievable with $\gamma=0$ [black solid line, panel (a)], and in terms of the variance of the nullifiers over $\rho'_{st}$, relative to the variance over the vacuum [dark gray lines, panels (b)].
We observe that significant reduction of the variance (squeezing) of the nullifiers 
(which indicates that the state is close to the cluster state) is observed when $\gamma\,\pt{N+1}\ll\kappa$, namely when the total added dissipation is much weaker than the dissipation in the squeezed reservoir.
The thin lines in panel (a) describe how the model is sensitive to deviation form the ideal Hamiltonian~\rp{H}. We have considered both deviation in the amplitude (thin solid gray lines) and in the phase  (thin dashed red lines) of the interaction coefficients, and we observe that the system is significantly more stable with respect to the latter. In any case, even when the fidelity is very low, the nullifiers always exhibit significant squeezing [panel (c)].

We note that the overlaps between normal modes and auxiliary mode [see panel (e)] determine the rates at which each normal mode is coupled to the squeezed reservoir. In the ideal case, 
these overlaps determine how fast each normal mode approaches 
the steady state. The optimal situation is the one in which all the overlaps are equal and are as large as possible so that all the normal modes are optimally coupled to the reservoir. This is described by Fig.~\ref{Fig2}, which shows that in this case the system is significantly more resistant to deviations from the ideal configuration.
We also note that the overlaps are the same for both $H\al{S}$ and $H$ (because $U\al{p}$ does not operate on the auxiliary mode~\cite{SM}). And this means that, for any state, the time to reach the steady state is entirely determined by the dynamics of the linear chain [Eq.~\rp{HSlinearChain}].

%
In conclusion, we have shown that, by squeezing the local environment of a single site of an harmonic lattice,
it is possible to steer the whole system toward any pure Gaussian state that can be generated
by a passive multi-mode transformation which operates on a batch of many equally squeezed modes.
In particular, given one of these states, we have shown how to determine a passive quadratic Hamiltonian which sustain it in the stationary regime (and which necessarily fulfills the chiral symmetry identified in Ref.~\cite{yanay2018}).
This Hamiltonian is not unique~\cite{SM}, and we have shown, by studying the generation of a cluster state in a square lattice, that the efficiency for the preparation of the chosen state, in non-ideal situations, depends critically on the specific ideal Hamiltonian that one considers. Understanding which Hamiltonian
is more suitable to its practical realization, and which Hamiltonian
 corresponds to a model which is more resistant to imperfections, are questions 
which deserve further investigation. Another interesting related question regards the possibility to extend this approach to spin systems~\cite{zippilli2013}.
Moreover, these findings also suggest how to extend the protocol discussed in~\cite{zippilli2013,zippilli2014} to entangle generic distant arrays using a two-mode squeezed field.

We finally note that this approach can be particularly valuable for implementations of quantum information devices
with circuit QED systems, which have been recently used to realize various lattice models~\cite{hacohen-gourgy2015,fitzpatrick2017,ma2019b}.
An experimental implementation of our results would require the ability to design the lattice Hamiltonian with one of these systems, and to combine it with a squeezed field of sufficiently large bandwidth~\cite{zippilli2014,asjad2016a}, produced for example with Josephson parametric amplifiers~\cite{castellanos-beltran2008,aumentado2020}.
Alternatively, the squeezed reservoir could be also engineered with bichromatic drives~\cite{zippilli2015}.

\acknowledgments{
We acknowledge the support of the European Union Horizon 2020 Programme for Research and Innovation through the Project No. 732894 (FET Proactive HOT), and the Project No. 862644 (FET Open QUARTET).}

\end{bibunit}



\onecolumngrid


\newpage

\thispagestyle{empty}

\setcounter{page}{1}
\setcounter{section}{0}
\setcounter{equation}{0}
\setcounter{figure}{0}
\renewcommand{\thesection}{S.\Roman{section}}
 \renewcommand{\theequation}{S.\arabic{equation}}
 \renewcommand{\thefigure}{S.\arabic{figure}}

\onecolumngrid

\hspace{-.5cm}
\begin{minipage}{18.cm}
\begin{center}
{\bf\large  Supplemental material for ``%
\makeatletter\@title\makeatother%
''
}

\bigskip 
{Stefano~Zippilli$^1$ and David~Vitali$^{1,2,3}$}
\vspace{0.1cm}

{\it \small
${}^1$School of Science and Technology, Physics Division, University of Camerino, 
	62032 Camerino (MC), Italy\\
${}^2$INFN, Sezione di Perugia, I-06123 Perugia, Italy\\
${}^3$CNR-INO, L.go Enrico Fermi 6, I-50125 Firenze, Italy
}

\smallskip
(Dated: \today)


\end{center}
\end{minipage}

\vspace{1cm}
\twocolumngrid

\begin{bibunit}

\section{Uniqueness of the steady state}

Here we show that the proposition~\ref{cond2b} of the main text guarantees that the system has no dark modes.
We consider the linear system of equations for the average annihilation operators 
$\av{\dot b_0}=-\frac{\ii}{\hbar}\av{\pq{b_0,H\al{S}}}-\kappa\,\av{b_0}$ and
$\av{\dot b_j}=-\frac{\ii}{\hbar}\av{\pq{b_j,H\al{S}}}$ for $j\neq0$,
which can be written in matrix form as 
$\av{\dot b_j}=-\sum_{k=0}^N\ \RR_{j,k}\,\av{b_k}$, for all $j$,
with $\RR=\ii\,\JJ\al{S}+\Gamma$, where $\JJ\al{S}$ is the hermitian matrix of coefficients corresponding to the Hamiltonian part and $\Gamma$ is the matrix for the dissipative part which has a single non-zero entry
\begin{eqnarray}
\Gamma=\pt{\mat{ccc}{
\kappa&0&\cdots\\
0&0&  \\
\vdots& & \ddots\\
}}\ .
\end{eqnarray}
First we note that when we say that our model has no dark modes, we mean that the matrix $\RR$ is positive stable~\cite{horn1991}, i.e.  all the eigenvalues of $\RR$ have positive real part, so that all the normal modes actually decay. 
According to the theorem 2.4.7 of Ref.~\cite{horn1991}, given a positive definite 
matrix $\QQ$, such that $\QQ\,\RR+\RR\da\,\QQ$ is semi-positive definite, then $\RR$ is positive stable if and only if no eigenvector of $\QQ^{-1}\,\pt{\QQ\,\RR-\RR\da\,\QQ}$ lies in the null space of $\QQ\,\RR+\RR\da\,\QQ$. In our case, we can simply choose $\QQ=\id$ (the identity matrix), such that $\QQ\,\RR+\RR\da\,\QQ=2\,\Gamma$, which is semi-positive definite, and $\QQ^{-1}\,\pt{\QQ\,\RR-\RR\da\,\QQ}=2\,\ii\,\JJ\al{S}$.
The subspace orthogonal to the null space of $\Gamma$ is given by the single vector $\vv_0=\pt{1,0,\cdots 0}^T$, corresponding to the auxiliary mode. Hence, when all the eigenvectors $\vw_j$ of $\JJ\al{S}$ are not orthogonal to $\vv_0$, i.e. the scalar product $\vw_j\cdot\,\vv_0\neq 0$ for all $j$ [which is equivalent to the proposition~\ref{cond2b} of the main text], then the conditions of the theorem are fulfilled, so that $\RR$ is positive stable.

\section{Gaussian States and the Bloch-Messiah decomposition}
\label{GBM}

In this work we consider $N$-modes, pure Gaussian states, which have zero average (no displacement). These states are given by
\begin{eqnarray}
\ke{\Psi}=U\,\ke{0}
\end{eqnarray}
where $\ke{0}$ is the vacuum and $U$ is a unitary transformation which can be expressed in terms of the vector of bosonic operator $\vb=\pt{b_1\cdots,b_N,b_1\da\cdots,b_N\da}^T$ and a $2\,N\times2\,N$ complex symmetric matrix $\SSS=\SSS^T$ as
\begin{eqnarray}\label{UeS}
U=\ee^{-\frac{\ii}{2}\,\vb^T\ \SSS\ \vb}\ .
\end{eqnarray}
The term $\vb^T\ \SSS\ \vb$ is Hermitian. This entails that the matrix $\SSS$ fulfills the relation $\SSS=\GG\,\SSS^*\,\GG$, with $\GG=\pt{\mat{cc}{&\id\\ \id&}}$ where the missing blocks are null matrices, and $\SSS^*$ is the matrix whose entries are the complex conjugates of the entries of $\SSS$. This means that $\SSS$ has the block structure
$\SSS=\pt{\mat{cc}{\ZZ^*&\KK^*\\ \KK&\ZZ}}$, where $\ZZ$ and $\KK$ are $N\times N$ complex matrices which, since $\SSS$ is symmetric, fulfill the relations $\ZZ=\ZZ^T$ and $\KK=\KK\da$.
 
The mode operators are transformed by the unitary $U$ according to a Bogoliubov matrix $\BB$ such as 
\begin{eqnarray}\label{UbUBb}
U\da\,\vb\,U=\BB\,\vb\ .
\end{eqnarray}
The matrix $\BB$ can be expressed in terms of the matrix $\SSS$ as
\begin{eqnarray}\label{BiJS}
\BB=\ee^{-\ii\,\II\,\SSS}
\end{eqnarray}
where 
$$\II=\pt{\mat{cc}{&\id\\ -\id&}}\ .$$ 
This can be shown using the Baker-Hausdorff formula ($\ee^{A}B\ee^{-A}=\sum_{n=0}^\infty\frac{1}{n!}\pq{A,B}\al{n}$ with $\pq{A,B}\al{0}=B$ and $\pq{A,B}\al{n}=\pq{A,\pq{A,B}\al{n-1}}$), such that
\begin{eqnarray}
\pg{\BB\,\vb}_j&=&
U\da\,\vb_j \,U
=\vb_j+\sum_{n=1}^\infty\frac{\ii^n}{2^n\,n!}\pq{\sum_{k,k'}\SSS_{k,k'}\,\vb_k\,\vb_{k'},\vb_j}\al{n}\ ,
\end{eqnarray}
where $\pq{\cdots,\cdots}\al{n}$ indicates the $n-$fold commutator such that $\pq{A,B}\al{1}=\pq{A,B}$, $\pq{A,B}\al{2}=\pq{A,\pq{A,B}}$, $\pq{A,B}\al{3}=\pq{A,\pq{A,\pq{A,B}}}$ and so on.
It is easy to show by induction, and using the bosonic commutation relations $\pq{\vb_j,\vb_k}=\II_{j,k}$, that   
\begin{eqnarray}
&&\pq{\sum_{k,k'}\SSS_{k,k'}\,\vb_k\,\vb_{k'},\vb_j}\al{n}
=\pt{-2}^n\pg{\pt{\II\SSS}^n\,\vb}_j
\end{eqnarray}
so that 
\begin{eqnarray}\label{BbUbU}
\pg{\BB\,\vb}_j&=&\pg{\pq{1+\sum_{n=1}^\infty\frac{\pt{-\ii}^n}{n!}\pt{\II\,\SSS}^n}\, \vb}_j\ .
\end{eqnarray}

The Bogoliubov matrix fulfills the relation $\BB=\GG\,\BB^*\GG$, and it can be expressed in block form as 
\begin{eqnarray}\label{BB}
\BB=\pt{\mat{cc}{\XX&\YY\\ \YY^*&\XX^*}} 
\end{eqnarray}
for some complex $N\times N$ matrices $\XX$ and $\YY$.
Moreover, due to the standard bosonic commutation relation for the transformed operators, $\BB$ fulfill also the relation $\BB\,\II\,\BB^T=\II$ which can be expressed in terms of the matrices $\XX$ and $\YY$ as
\begin{eqnarray}\label{BogCond1}
\XX\,\XX\da-\YY\,\YY\da=\id
\end{eqnarray}
and 
\begin{eqnarray}\label{BogCond2}
\XX\,\YY^T=\YY\,\XX^T \ .
\end{eqnarray}

The Block-Messiah~\cite{braunstein2005,vanloock2007,gu2009,cariolaro2016a,cariolaro2016} reduction formula allows to decompose $\BB$ as the product of three Bogoliubov transformations 
\begin{eqnarray}\label{BM0}
\BB=
\matt{\VV^\circ&}{&{\VV^\circ}^*}\,\matt{\DD_x &\DD_y^\circ}{\DD_y^\circ&\DD_x}\, \matt{{\WW^\circ}\da&}{&{\WW^\circ}^T}
\end{eqnarray}
where $\DD_x$ and $\DD_y^\circ$ are semi-positive definite
diagonal matrices and $\VV^\circ$ and $\WW^\circ$ are unitary matrices. They correspond to the singular value decomposition of the matrices $\XX$ and $\YY$, such that
\begin{eqnarray}\label{B-M}
\XX&=&\VV^\circ\,\DD_x\,{\WW^\circ}\da
\nn\\
\YY&=&\VV^\circ\,\DD_y^\circ\,{\WW^\circ}^T\ ,
\end{eqnarray}
and the diagonal elements of $\DD_x$ and $\DD_y^\circ$ are the singular values of $\XX$ and $\YY$ respectively.
The first and third matrices in Eq.~\rp{BM0} describe passive multi-mode transformations, while the second one describes the single-mode-squeezing of all the modes.
It is possible to include generic squeezing phases to the second transformation by defining these three matrices
\begin{eqnarray}
\VV\ &=&\ \VV^\circ\ \ee^{-\frac{\ii}{2}\,\Phi} 
\nn\\
\WW\ &=&\ \WW^\circ \ee^{-\frac{\ii}{2}\,\Phi}\ 
\nn\\
\DD_y\ &=&\ \DD_y^\circ\ \ee^{\ii\,\Phi} 
\end{eqnarray}
where $\Phi$ is a real diagonal matrix, and where now $\DD_y$ have complex entries, such that we can write this decomposition
\begin{eqnarray}\label{BM}
\BB=\BB_\VV\ \BB_\DD\ \BB_\WW
\end{eqnarray}
with
\begin{eqnarray}
\BB_\VV&=&\matt{\VV&}{&\VV^*}\ ,
\nn\\
\BB_\DD&=&\matt{\DD_x &\DD_y}{\DD_y^*&\DD_x}\ ,
\nn\\
\BB_\WW&=&\matt{\WW\da&}{&\WW^T}\ .
\end{eqnarray}
These three matrices correspond to three unitary transformations 
\begin{eqnarray}\label{UVUDUW}
U_\VV&=&\ee^{-\frac{\ii}{2}\,\vb^T\,\SSS_\VV\,\vb}\ ,
\nn\\
U_\DD&=&\ee^{-\frac{\ii}{2}\,\vb^T\,\SSS_\DD\,\vb}\ ,
\nn\\
U_\WW&=&\ee^{-\frac{\ii}{2}\,\vb^T\,\SSS_\WW\,\vb}\ ,
\end{eqnarray}
(for some matrices $\SSS_\VV$, $\SSS_\DD$ and $\SSS_\WW$ which are specified below) which can be used to decompose the $U$ as 
\begin{eqnarray}
U=U_\VV\ U_\DD\ U_\WW\ ,
\end{eqnarray}
such that $U\da\,\vb\,U=\BB\,\vb=\BB_\VV\,\BB_\DD\,\BB_\WW\,\vb=\BB_\VV\,\BB_\DD\ U_\WW\da\,\vb\,U_\WW=U_\WW\da\pt{\BB_\VV\,\BB_\DD\,\vb}\,U_\WW=U_\WW\da\,U_\DD\da\,U_\VV\da\ \vb\ U_\VV\ U_\DD\ U_\WW$. 
By means of these operators we find that a general $N$-modes, zero-average, pure Gaussian states can be expressed as 
\begin{eqnarray}\label{PsiUVUD0}
\ke{\Psi}=U_\VV\ U_\DD\ke{0}\ ,
\end{eqnarray}
where the vacuum is not changed by the passive transformation $U_\WW$. 
This corresponds to the decomposition introduced in the main text with
\begin{eqnarray}\label{USUDUpUV}
U\al{S}&=&U_\DD
\nn\\
U\al{p}&=&U_\VV\ .
\end{eqnarray}

Since $\VV$ is a unitary matrix, it can be expressed as
\begin{eqnarray}
\VV&=&\ee^{-\ii\,\KK_v}\ ,
\end{eqnarray}
for a $N\times N$ hermitian matrix $\KK_v$
so that 
\begin{eqnarray}\label{SSSV}
\SSS_\VV&=&\matt{&\KK_v^*}{\KK_v&}\  
\end{eqnarray}
(similar considerations hold also for $\SSS_\WW$).
Moreover the matrices  $\DD_x$ and $\DD_y$ are diagonal and have to fulfill a condition analogous to Eq.~\rp{BogCond1}. This implies that they can be rewritten in terms of a diagonal matrix 
$\DD_z$ as 
\begin{eqnarray}
\DD_x&=&\cosh\pt{\DD_z}\ ,
\nn\\
\DD_y&=&\sinh\pt{\DD_z}\ \ee^{\ii\,\Phi}\ ,
\end{eqnarray}
and, in turn, the corresponding unitary transformation $U_\DD$ [see Eq.~\rp{UVUDUW}], which describes a batch of single-mode squeezing transformations, is expressed in terms of the matrix 
\begin{eqnarray}
\SSS_\DD=\matt{-\ii\,\DD_z\ \ee^{-\ii\,\Phi}&}{&\ii\,\DD_z\ \ee^{\ii\,\Phi}}\ ,
\end{eqnarray}
where the non-zero entries of the the diagonal matrix $\Phi$ are $\Phi_{j,j}=\varphi_j$, with $\varphi_j$ the squeezing phases introduced in the main text.

In the main text we have shown that with our approach it is possible to generate any state of the form~\rp{PsiUVUD0} where $U_\DD=U\al{S}$  describes the equal squeezing for all the modes,
namely states for which $\DD_z=z\,\id$
for some real non-negative $z$, so that
\begin{eqnarray}\label{SSSD}
\SSS_\DD=\matt{-\ii\,z\ \ee^{-\ii\,\Phi}\ \id&}{&\ii\,z\ \ee^{\ii\,\Phi}\ \id}\ .
\end{eqnarray}
In other terms we can prepare states for which the singular values of the blocks that constitute the corresponding Bogoliubov transformation, $\XX$ and $\YY$, are all equal, i.e. $\DD_x=\cosh(z)\,\id$ and $\DD_y^\circ=\sinh(z)\,\id$. Since the singular values of a generic matrix $\MM$ are the square roots of the eigenvalues of $\MM\ \MM\da$, this means that the matrices $\XX$ and $\YY$ are proportional to unitary matrices, i.e. $\XX\ \XX\da=\cosh^2(z)\ \id$ and $\YY\ \YY\da=\sinh^2(z)\ \id$.

\section{Proof of the Lemma of the main text}

It is straightforward to prove the lemma by noting that, on the one hand, on site energy terms $b_j\da\,b_j$ result in non-passive single mode squeezing terms under the transformation $U_j$, and that, on the other hand,
interaction terms $h_{j,k}=\hbar\,\pt{\JJ_{j,k}\al{S}\,b_j\da\,b_k+{\JJ_{j,k}\al{S}}^*\,b_k\da\,b_j}$, with $j\neq k$, are invariant under the effect of the transformation $U_j\,U_k$, namely ${U_k}\da\,{U_j}\da\ h_{j,k}\ U_j\,U_k=h_{j,k}$, if and only if the proposition (ii) is true; different squeezing strengths or phases result instead in non-passive two-mode squeezing terms in the transformed Hamiltonian.

To be specific, Given the squeezing operator $U_j=\ee^{\frac{z_j}{2}\pt{\ee^{\ii\varphi_j}{b_j\da}^2-\ee^{-\ii\varphi_j}{b_j}^2}}$
we find 
$U_j\da\ b_j\ U_j=c_j\,b_j+s_j\,\ee^{\ii\,\varphi_j}\,b_j\da$, with $c_j=\cosh(z_j)$ and $s_j=\sinh(z_j)$. Thus,
given the Hamiltonian $H\al{S}$, which we rewrite as
$H\al{S}=
\hbar\,\sum_{j=0}^N\,\JJ_{j,j}\al{S}\,b_j\da\,b_j+
\hbar\,\sum_{j<k=0}^N\,\pt{
\JJ_{j,k}\al{S}\,b_j\da\,b_k+{\JJ_{j,k}\al{S}}^*\,b_k\da\,b_j
}$
we find
\begin{eqnarray}
\wt H&=&U_N\da\,\cdots U_0\da\ H\al{S}\ U_0\cdots U_N
\nn\\
&=&\hbar\,\sum_{j=0}^N\,\JJ_{j,j}\al{S}\pq{
c_j^2\,b_j\da\,b_j+s_j^2\,b_j\,b_j\da+c_j\,s_j\pt{\ee^{\ii\,\varphi_j}\,{b_j\da}^2+\ee^{-\ii\,\varphi_j}\,b_j^2}
}
\nn\\&&
+\hbar\,\sum_{j<k=0}^N\,\lpg{
\pq{\JJ_{j,k}\al{S}\,c_j\,c_k+{\JJ_{j,k}\al{S}}^*\,s_j\,s_k\,\ee^{\ii(\varphi_j-\varphi_k)}}b_j\da\,b_k
}
\nn\\&&\rpg{
+\pq{\JJ_{j,k}\al{S}\,c_j\,s_k\,\ee^{\ii\,\varphi_k}+{\JJ_{j,k}\al{S}}^*\,s_j\,c_k\,\ee^{\ii\,\varphi_j}}b_j\da\,b_k\da
+h.c.}\ .
\end{eqnarray}
This Hamiltonian is passive if and only if
\begin{eqnarray}\label{SM:lemma1}
\JJ_{j,j}\al{S}\,c_j\,s_j&=&0
\\
\JJ_{j,k}\al{S}\,c_j\,s_k\,\ee^{\ii\,\varphi_k}+{\JJ_{j,k}\al{S}}^*\,s_j\,c_k\,\ee^{\ii\,\varphi_j}&=&0
\label{SM:lemma2}
\end{eqnarray}
for all $j<k$. Finally, we note that, Eq.~\rp{SM:lemma1} is equivalent to the proposition (i) of the lemma, and Eq.~\rp{SM:lemma2} is equivalent to $\frac{s_j\,c_k}{c_j\,s_k}=-
\frac{\JJ_{j,k}\al{S}}{{\JJ_{j,k}\al{S}}^*}
\ \ee^{\ii\,(\varphi_k-\varphi_j)}$, which
is equivalent to the proposition (ii) of the lemma. In particular, in this case 
\begin{eqnarray}
\wt H&=&\hbar\,\sum_{j<k=0}^N\,\pg{
\pq{\JJ_{j,k}\al{S}\,c_j^2+{\JJ_{j,k}\al{S}}^*\,s_j^2\,\ee^{\ii(\varphi_j-\varphi_k)}}b_j\da\,b_k
+h.c.}
\nn\\
&=&H\al{S}\ .
\end{eqnarray}
$\hfill\blacksquare$

\section{Relation between the chiral symmetry of Ref.~\cite{yanay2018} and the present result}

\subsection{The chiral symmetry of Ref.~\cite{yanay2018} and the Hamiltonians $H\al{S}$ and $H$}

Here we show that the Hamiltonians $H\al{S}$ and $H$ of the main text satisfy the chiral symmetry discussed in Ref.~\cite{yanay2018}.

According to the lemma of the main text, the Hamiltonian $H\al{S}$
can be expressed as
\begin{eqnarray}\label{SM:HS}
H\al{S}=\hbar\,\sum_{j,k=0}^N \,\JJ_{j,k}\al{S}\ b_j\da\,b_k
\end{eqnarray}
where $\JJ\al{S}$ is a $(N+1)\times(N+1)$ Hermitian matrix with entries $\JJ_{j,j}\al{S}=0$ and $\JJ_{j,k}\al{S}=\ii\,\abs{\JJ_{j,k}\al{S}}\ \ee^{\ii\pt{\varphi_j-\varphi_k}/2}$, for $j<k$.
It can be decomposed as
$\JJ\al{S}=\ee^{\ii\,\Phi}\,\wt{\JJ\al{S}}\,\ee^{-\ii\,\Phi}$, 
where $\Phi$ is the diagonal matrix with entries $\Phi_{j,j}=\varphi_j/2$, and $\wt{\JJ\al{S}}$ is an Hermitian matrix with imaginary entries. The matrices $\JJ\al{S}$ and $\wt{\JJ\al{S}}$ have the same eigenvalues $\lambda_j$ and the eigenvectors $\vw_j$ of $\JJ\al{S}$ are related to the eigenvectors $\wt\vw_j$ of $\wt{\JJ\al{S}}$ by the relation $\vw_j=\ee^{\ii\,\Phi}\,\wt\vw_j$. 
Given an eigenvalue $\lambda_j$ and the corresponding eigenvector $\wt\vw_j$, if we take the complex conjugate of $\wt{\JJ\al{S}}\ \wt\vw_j=\lambda_j\ \wt\vw_j$, we find that (since $\wt{\JJ\al{S}}$ is imaginary) $-\lambda_j$
is the eigenvalue corresponding to the eigenvector $\wt\vw_j^*$. And finally, this means that $H\al{S}$ fulfills the chiral symmetry  of Ref.~\cite{yanay2018}. Namely, the normal modes of $H\al{S}$ [i.e. the eigenvectors of $\JJ\al{S}$ and $\wt{\JJ\al{S}}$] come in pairs, with opposite frequencies, such that, by proper reordering of the normal modes, $\lambda_j=-\lambda_{j+1}$ (for and odd number of modes there is also a zero-frequency mode, $\lambda_0=0$); and, moreover, the overlap between the auxiliary mode, described by the vector $\vv_0=\pt{1,0,\cdots 0}^T$, and the normal mode $\vw_j$ is equal in modulus to the overlap between $\vv_0$ and the normal mode with opposite frequency $\vw_{j+1}$ (which, as we have seen, is given by $\vw_{j+1}=\vw_j^*$), such that ${\vv_0\cdot\vw_j}=\pt{\vv_0\cdot\vw_{j+1}}^*$.

Correspondingly, the passive Hamiltonian $H$
of the model~\rp{MEq}
of the main text can be expressed in terms of a $(N+1)\times(N+1)$ Hermitian matrix $\JJ$, as 
\begin{eqnarray}\label{SM:H}
H=\hbar\,\sum_{j,k=0}^N \,\JJ_{j,k}\ b_j\da\,b_k\ .
\end{eqnarray}	
It is related to $H\al{S}$ by the unitary passive transformation $U\al{p}$ [see the
proposition~\ref{cond2} of the theorem
of the main text], which does not act on the auxiliary mode, and which can be expressed in terms of a $N\times N$ Hermitian matrix $\KK\al{p}$ as $U\al{p}=\ee^{-\ii\,\sum_{j,k=1}^N\ \KK_{j,k}\al{p}\ b_j\da\,b_k}$. 
Therefore the matrices $\JJ\al{S}$ [see Eq.~\rp{SM:HS}] and $\JJ$ [see Eq.~\rp{SM:H}] are related by a $(N+1)\times(N+1)$ unitary matrix $\UU$,
according to
\begin{eqnarray}
\JJ=\UU\da\ \JJ\al{S}\ \UU \ ,
\end{eqnarray}
where $\UU$ can be constructed in terms of the $N\times N$ matrix $\KK\al{p}$ which enters into the definition of $U\al{p}$ as
\begin{eqnarray}
\UU=\pt{
\mat{cc}{
1 & \cdots \\ \vdots & \ee^{-\ii\,\KK\al{p}}
}
}\ ,
\end{eqnarray} 
where the missing entries are all zeros.
This means that, on the one hand, the spectrum of $\JJ$ is equal to the spectrum of $\JJ\al{S}$, and that, on the other, given an eigenvector $\vw_j$ of $\JJ\al{S}$, the corresponding eigenvector of $\JJ$ is $\UU\da\,\vw_j$. In particular we find that the overlap with the auxiliary mode is equal for the eigenvectors of $\JJ\al{S}$ and for the corresponding eigenvectors of $\JJ$, i.e. $\vv_0\cdot\UU\da\,\vw_j=\vv_0\cdot\vw_j$. And, in turn, this entails that also $H$, fulfills the chiral symmetry of Ref.~\cite{yanay2018}.

\subsection{The chiral symmetry of Ref.~\cite{yanay2018} and the transformation which generates the steady state}

Here we show that the passive transformation $U\al{p}$ (defined in the theorem of the main text) which is part of the transformation which generates the steady state, is related to the passive unitary transformation $U\al{\TT}$ which diagonalize the Hamiltonian $H$ [such that ${U\al{\TT}}\da\,H\,U\al{\TT}=\hbar\sum_{j=0}^N\,\lambda_j\,b_j\da\,b_j$], according to the relation 
\begin{eqnarray}
U\al{p}=U\al{\TT}\ \wt U \ ,
\end{eqnarray}
where $\wt U$ is the product of many beam splitter interactions 
and phase shifts 
between the normal modes at opposite frequency,
the specific form of which is specified below.

This can be shown as follows. 
Let us, first, introduce the unitary matrix 
\begin{eqnarray}
\TT=\pq{\vw_0\ \vw_1\ \cdots\ \vw_N}\ , 
\end{eqnarray}
which diagonalize $\JJ$ (i.e. $\vw_j$ are the eigenvectors of $\JJ$ and $\JJ\,\TT=\TT\,\Lambda$, with $\Lambda$ the diagonal matrix with entries $\Lambda_{j,j}=\lambda_j$), and which can be expressed in terms of a hermitian matrix $\KK_\TT$ as 
\begin{eqnarray}
\TT=\ee^{-\ii\,\KK_\TT}\ . 
\end{eqnarray}
The density matrix 
$\rho\al{\TT}={U\al{\TT}}\da\ \rho\ U\al{\TT}$, with 
\begin{eqnarray}
U\al{\TT}=\ee^{-\ii\, \sum_{j,k=0}^N\,\KK_\TT\,b_j\da\,b_k}\ ,
\end{eqnarray}
fulfills the master equation 
$\dot\rho\al{\TT}=-\frac{\ii}{\hbar}\pq{H\al{\TT},\rho\al{\TT}}+\LL\al{\TT}\,\rho\al{\TT}$,
where $H\al{\TT}=\hbar\sum_{j=0}^N\,\lambda_j\,b_j\da\,b_j$ and $\LL\al{\TT}\ \rho\al{\TT}={U\al{\TT}}\da\,\pq{\LL\,\pt{U\al{\TT}\,\rho\al{\TT}\,{U\al{\TT}}\da} }U\al{\TT}$. 
It has been shown in Ref.~\cite{yanay2018} that 
in this representation, the steady state is characterized by many two-mode squeezed pairs, corresponding to the normal modes with opposite frequency $\lambda_j$ and $\lambda_{j+1}=-\lambda_j$ (in the case of an odd number of modes, the mode with zero frequency, that is the one with index $j=0$, is in a single-mode squeezed state), generated by the transformations
\begin{eqnarray}
U_{j}\al{2}=\ee^{-\ii\,z_0\pt{\ee^{\ii\,\varphi_0}\,b_{j}\da\,b_{j+1}\da+\ee^{-\ii\,\varphi_0}\,b_{j}\,b_{j+1}}}
\end{eqnarray}
(with $\varphi_0$ the phase of the squeezed reservoir defined in the main text),
where for an even number of modes ($N$ odd) $j$ takes even values $j\in\pg{0,2,4\cdots(N-1)}$,
instead, for an odd number of modes ($N$ even) $j$ takes odd values $j\in\pg{1,3,5\cdots N-1}$.
Thus, we introduce the transformation which generates all the entangled pairs, that is $U\al{TMS}=U_0\al{2}\,U_2\al{2}\,U_4\al{2}\cdots$ for and even number of modes, and $U\al{TMS}=U_0\,U_1\al{2}\,U_3\al{2}\cdots$ for and odd number of modes (where $U_0$ is the single mode squeezing transformation  defined in the main text), and we find that, in this representation, the steady state is $U\al{TMS}\,\ke{0}$. Correspondingly, in the original representation, the steady state can be written as
\begin{eqnarray}
\ke{\Psi_{tot}}=U\al{\TT}\,U\al{TMS}\,\ke{0}\ .
\end{eqnarray} 

Let us, now, consider the 50/50 beam splitter transformations 
between all the entangled pairs 
\begin{eqnarray}
U_j\al{BS}=\ee^{-\ii\,\frac{\pi}{4}\,\pt{b_{j}\da\,b_{j+1}+b_{j+1}\da\,b_{j}}}\ ,
\end{eqnarray}
[where $j$ is even (odd) for an even (odd) number of modes]
which realizes the transformations 
\begin{eqnarray}
{U_j\al{BS}}\da\ b_{j}\ U_j\al{BS}&=&\frac{1}{\sqrt{2}}\pt{b_{j}-\ii\,b_{j+1}}
\nn\\
{U_j\al{BS}}\da\ b_{j+1}\ U_j\al{BS}&=&\frac{1}{\sqrt{2}}\pt{-\ii\,b_{j}+b_{j+1}}
\nn\\
\end{eqnarray}
and the phase shifts for all the modes
\begin{eqnarray}
U_j\al{\phi}&=&\ee^{-\ii\,\xi_{j}\,b_{j}\da\,b_{j}}\ \ee^{-\ii\,\xi_{j+1}\,b_{j+1}\da\,b_{j+1}}
\end{eqnarray}
with
\begin{eqnarray}
\xi_{j}&=&\frac{\varphi_{j}-\varphi_0}{2}
\end{eqnarray}
(where $\varphi_j$ are the squeezing phases introduced in the proposition~\ref{cond1} of the theorem of the main text), which realizes the transformations
\begin{eqnarray}
{U_j\al{\phi}}\da\ b_{j}\ U_j\al{\phi}&=&b_{j}\,\ee^{-\ii\,\xi_{j}}
\nn\\
{U_j\al{\phi}}\da\ b_{j+1}\ U_j\al{\phi}&=&b_{j+1}\,\ee^{-\ii\,\xi_{j+1}}\ .
\end{eqnarray}
We find
\begin{eqnarray}
{U_j\al{\phi}}\da\ {U_j\al{BS}}\da\ \ U_j\al{2}\ U_j\al{BS}\ U_j\al{\phi} = U_{j}\ U_{j+1}
\end{eqnarray}
(with $U_j$ the single-mode squeezing transformation defined in the proposition~\ref{cond1} of the theorem of the main text).
So, if we define the passive unitary transformation
\begin{eqnarray}
\wt U&=&U_0\al{\phi}\ U_0\al{BS}\ U_2\al{\phi}\ U_2\al{BS}\cdots
\nn\\
(\wt U&=&U_1\al{\phi}\ U_1\al{BS}\ U_3\al{\phi}\ U_3\al{BS}\cdots)
\end{eqnarray}
for an even (odd) number of modes, we find
\begin{eqnarray}\label{UUTMSUUS}
\wt U\da\, U\al{TMS}\, \wt U=U\al{S}\,U_0 \ ,
\end{eqnarray}
where $U\al{S}$ is defined in the main text,
and therefore
\begin{eqnarray}
\ke{\Psi_{tot}}&=&U\al{\TT}\,\wt U\ \wt U\da\,U\al{TMS}\,\wt U\ \wt U\da\ke{0}
\nn\\
&=&U\al{\TT}\,\wt U\ U\al{S}\,U_0\,\ke{0}
\ ,
\end{eqnarray} 
where in the last step we have used Eq.~\rp{UUTMSUUS} and the fact that $\wt U\da$ is a passive transformation that does not change the vacuum.
In the main text, instead, we have shown that $\ke{\Psi_{tot}}=U\al{p}\,U\al{S}\,U_0\,\ke{0}$, and therefore 
\begin{eqnarray}
U\al{p}=U\al{\TT}\,\wt U\ .
\end{eqnarray}

\begin{figure}[t!]
\centering
\includegraphics[width=0.5\columnwidth]{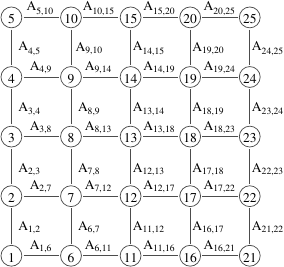}
\caption{Graph corresponding to the symmetric adjacency matrix $\AAA$ of the cluster state of $N=25$ modes in a square lattice discussed in the main text. The non-zero entries of $\AAA$ are equal to one and correspond to the edges of the graph.}
\label{SM:Fig0}
\end{figure}
\begin{figure}[t!]
\centering
\includegraphics[width=\columnwidth]{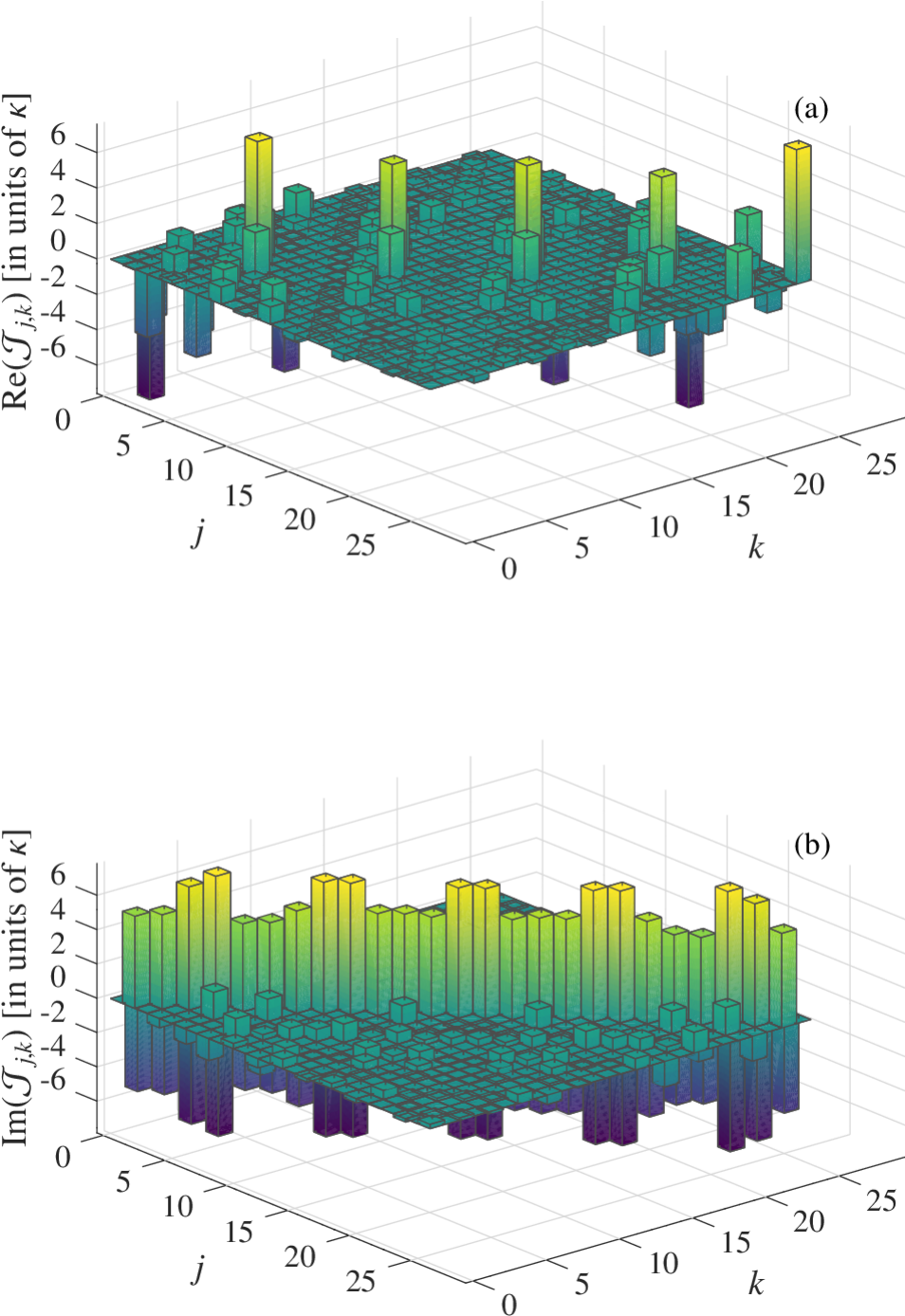}
\caption{(a) Real and (b) Imaginary parts of the interaction coefficients $\JJ_{j,k}$ of the Hamiltonian~\rp{H} of the main text used for the results of Fig.~\ref{Fig1} of the main text.}
\label{SM:Fig1}
\end{figure}
\begin{figure}[t!]
\centering
\includegraphics[width=\columnwidth]{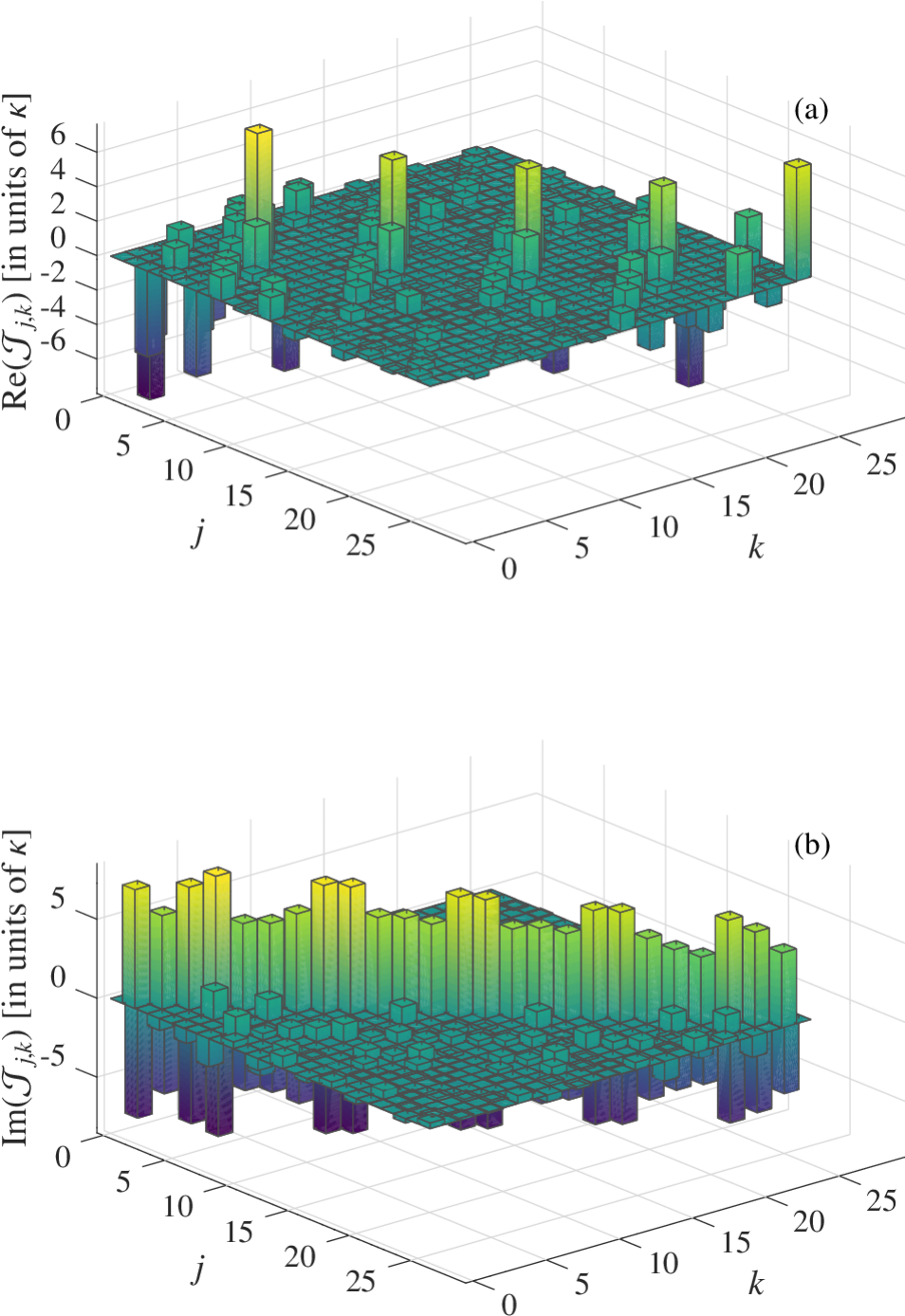}
\caption{(a) Real and (b) Imaginary parts of the interaction coefficients $\JJ_{j,k}$ of the Hamiltonian~\rp{H} of the main text used for the results of Fig.~\ref{Fig2} of the main text.}
\label{SM:Fig2}
\end{figure}

\section{Cluster states which can be prepared with the present approach}\label{cluster}

Given a $N\times N$ real symmetric adjacency  matrix $\AAA$, 
the corresponding cluster states is the zero eigenstate of the collective operators (called nullifiers)
\begin{eqnarray}\label{nullifiers}
x_j&=&p_j-\sum_{k=1}^N\,\AAA_{j,k}\,q_k\ ,\ \ \ \ \ \  
{\rm for}\ j\in\pg{1\cdots,N}\ ,
\end{eqnarray}
with $p_j=-\ii\pt{b_j-b_j\da}$ and $q_j=b_j+b_j\da$.
In other terms, these collective quadratures are infinitely squeezed for a cluster state with adjacency matrix $\AAA$. To be specific a cluster state can be written as $\ke{\Psi_{cluster}}=\ee^{\frac{\ii}{4}\sum_{j,k}\,A_{j,k}\,q_j\,q_k}\,\ke{0}_p$, where $\ke{0}_p$ is the infinitely squeezed states that is the zero eigenstate of the operators $p_j$, i.e $p_j\ke{0}_p=0$ $\forall\ j$. 

For realistic, approximated cluster states, this state is squeezed by a finite amount.
In general an approximated cluster state $\ke{\Psi_z}$ can be defined in terms of a finite squeezing parameter $z$ and the adjacency matrix $\AAA$,  such that the covariance matrix of the nullifiers $\CC_z=\br{\Psi_z}\,{\vx\vx^T+\pt{\vx\vx^T}^T}\,\ke{\Psi_z}/2$ approaches the null matrix in the limit $z\to\infty$.

An example is given by the state generated by the unitary transformation
\begin{eqnarray}\label{Uz1}
U_z=\ee^{\frac{i}{4}\sum_{j,k}\,\AAA_{j,k}\ q_j\,q_k}\ \ee^{\frac{z}{2}\sum_j\pt{b_j\da{}^2-b_j^2}}\ ,
\end{eqnarray}
that is $\ke{\Psi_{z}}=U_z\,\ke{0}$, where $\ke{0}$ is the vacuum.
In this case we find that the corresponding Bogoliubov matrix 
has the structure of Eq.~\rp{BB} with
\begin{eqnarray}\label{XXzYYz-a}
\XX_z&=&\id\ \cosh(z)+\frac{\ii}{2}\,\ee^{z}\ \AAA
\nn\\
\YY_z&=&\id\ \sinh(z)+\frac{\ii}{2}\,\ee^{z}\ \AAA\ .
\end{eqnarray}
It is possible to check that the covariance matrix of the nullifiers~\rp{nullifiers} approaches the null matrix in the limit of large $z$. To be specific in this case $\CC_z=\ee^{-2\,z}\,\id$.
In our approach we can construct states for which the singular values of the matrices~\rp{XXzYYz-a} are all equal. In other terms the matrices $\XX_z\,\XX_z\da=\id\ \cosh^2(z)+\frac{\ee^{2\,z}}{4}\,\AAA^2$ and $\YY_z\,\YY_z\da=\id\ \sinh^2(z)+\frac{\ee^{2\,z}}{4}\,\AAA^2$ have to be proportional to the identity. This implies that with our approach we can construct cluster states given by Eq.~\rp{Uz1} for which the adjacency matrix is proportional to a self-inverse matrix $\AAA^2=\alpha\,\id$ for some positive real $\alpha$.

Another example is given by a multi-mode squeezed state generated by the transformation 
\begin{eqnarray}\label{Uz2}
U_z=\ee^{-\ii\,\frac{z}{2}\ \vb^T\,\SSS_z\,\vb}
\end{eqnarray}
with
\begin{eqnarray}
\SSS_z=\pt{\mat{cc}{\ZZ^*&\\&\ZZ}}\ ,
\end{eqnarray}
where $\ZZ$ is a complex symmetric, non-singular matrix. 
In Ref.~\cite{zippilli2020} we have shown that these states are approximated cluster states, which can be realized using many equally squeezed modes, when the matrix $\ZZ$ is related to the adjacency matrix $\AAA$ by the relation
\begin{eqnarray}
\ZZ
&=&-\ii\,\frac{\AAA-\ii\,\id}{\AAA+\ii\,\id}\ ,
\end{eqnarray}
such that it is unitary.
In this case 
\begin{eqnarray}
\XX&=&\cosh(z)\ \id
\nn\\
\YY&=&-\ii\,\sin(z)\ \ZZ
\end{eqnarray}
and the Bloch-Messiah decomposition~\rp{BM} is given by
\begin{eqnarray}\label{DVW}
\DD_x&=&\cosh(z)\ \id
\nn\\
\DD_Y^\circ&=&\sinh(z)\ \id
\nn\\
\VV^\circ&=&\pt{-\ii\,\ZZ}^{1/2}\ 
\OO
\nn\\
\WW^\circ&=&\pt{-\ii\,\ZZ\ }^{1/2}\
\OO
\ ,
\end{eqnarray}
where $\OO$ is a generic real orthogonal matrix, and
where the last two matrices are found by the Autonne--Takagi factorization~\cite{horn2013,cariolaro2016a,cariolaro2016} of the symmetric unitary 
$-\ii\,\ZZ$, such that $-\ii\,\ZZ=\pt{-\ii\,\ZZ}^{1/2}\ {\pt{-\ii\,\ZZ}^{1/2}}^T$~\cite{zippilli2020}.
In the main text we have studied the preparation of a state of this form where the adjacency matrix $\AAA$ represents the square lattice depicted in Fig.~\ref{SM:Fig0}. The decomposition of the corresponding  unitary transformation $U_z=U\al{p}\,U\al{S}$ (see the proposition~\ref{cond1} in the theorem of the main text)  can be found as discussed in Sec.~\ref{GBM}. See in particular  Eqs.~\rp{UVUDUW},\rp{USUDUpUV}, \rp{SSSV} and \rp{SSSD}, where in this case the matrix $\VV^\circ$ is given in Eq.~\rp{DVW}. 
Note that the results of the main text are found with $\OO=\id$, and that a different $\OO$ corresponds to a different $U\al{p}$, and thus to a different system Hamiltonian $H$ of the main text.

In Figs.~\ref{SM:Fig1} and \ref{SM:Fig2} we report the coefficients of the system Hamiltonians that we have used in the result presented in the main text. In particular, in the main text, we have shown that the steady state of Eq.~\rp{MEqtot} of the main text, with the Hamiltonian represented in Figs.~\ref{SM:Fig1} and \ref{SM:Fig2}, approximates the cluster state with the adjacency matrix represented in Fig.~\ref{SM:Fig0}.

\end{bibunit}

\end{document}